\documentclass[Journal]{IEEEtran}
\IEEEoverridecommandlockouts
\usepackage{cite}
\usepackage{float}
\usepackage{amsmath,amssymb,amsfonts}
\usepackage[ruled,linesnumbered,vlined]{algorithm2e}
\usepackage{algpseudocode}
\usepackage{graphicx}
\usepackage{textcomp}
\usepackage{xcolor}
\usepackage{multirow}
\usepackage{diagbox}
\usepackage{array}
\usepackage{footnote}
\makesavenoteenv{tabular}
\makesavenoteenv{table}
\begin{document}

\title{Adaptive Target-Condition Neural Network: DNN-Aided Load Balancing for Hybrid LiFi and WiFi Networks}

\author{Han Ji, \textit{Student Member, IEEE}, Qiang Wang, \textit{Student Member, IEEE}, Stephen J. Redmond, \textit{Senior Member, IEEE}, Iman Tavakkolnia, \textit{Senior Member, IEEE}, and Xiping Wu, \textit{Senior Member, IEEE}

\thanks{The work of H. Ji and Q. Wang was supported by the China Scholarship Council (Grant No. 202106620012 and 202006540003). S. J. Redmond is partly supported by a Science Foundation Ireland President of Ireland Future Research Leaders Award (17/FRL/4832). X. Wu acknowledges the support of Royal Irish Academy (RIA) Charlemont Grant.}

\thanks{H. Ji, Q. Wang, S. J. Redmond, and X. Wu are with the School of Electrical and Electronic Engineering, University College Dublin, Belfield, Dublin, D04 V1W8, Ireland (e-mail:\{han.ji, qiang.wang\}@ucdconnect.ie; \{stephen.redmond, xiping.wu\}@ucd.ie).}

\thanks{I. Tavakkolnia is with the Electronic and Electrical Engineering Department, University of Strathclyde, Glasgow, G1 1XQ, United Kingdom (e-mail: i.tavakkolnia@strath.ac.uk).}

\thanks{\textit{Corresponding author: Xiping Wu}.}
}
\markboth{Journal of \LaTeX\ Class Files,~Vol.~XX, No.~XX, July~2022}%
{How to Use the IEEEtran \LaTeX \ Templates}

\maketitle
\begin{abstract}
Load balancing (LB) is a challenging issue in the hybrid light fidelity (LiFi) and wireless fidelity (WiFi) networks (HLWNets), due to the nature of heterogeneous access points (APs). Machine learning has the potential to provide a complexity-friendly LB solution with near-optimal network performance, at the cost of a training process. The state-of-the-art (SOTA) learning-aided LB methods, however, need retraining when the network environment (especially the number of users) changes, significantly limiting its practicability. In this paper, a novel deep neural network (DNN) structure named adaptive target-condition neural network (A-TCNN) is proposed, which conducts AP selection for one target user upon the condition of other users. Also, an adaptive mechanism is developed to map a smaller number of users to a larger number through splitting their data rate requirements, without affecting the AP selection result for the target user. This enables the proposed method to handle different numbers of users without the need for retraining. Results show that A-TCNN achieves a network throughput very close to that of the testing dataset, with a gap less than 3\%. It is also proven that A-TCNN can obtain a network throughput comparable to two SOTA benchmarks, while reducing the runtime by up to three orders of magnitude.

\end{abstract}


\section{Introduction}
The Cisco Annual Internet Report (2018-2023) forecasts a four-fold increase in the number of wireless fidelity (WiFi) hotspots between 2018 and 2023, to 628 million by the end of 2023 [\ref{Ref: Cisco White Paper}]. The dense deployment of WiFi hotspots would lead to intense competitions for available channels, due to the limited radio-frequency (RF) spectrum. This drives research on light fidelity (LiFi) [\ref{Ref: What is LiFi}], which operates in a way similar to WiFi but explores the extremely wide visible light spectrum ($\sim$300 THz). In contrast to WiFi, LiFi offers several prominent advantages including licence-free, availability in RF-restricted areas, and high security. Upon the existing light infrastructure, LiFi can also realize illumination and communication simultaneously, boosting the energy efficiency. Recent experimental results showed that LiFi is capable of rendering a link data rate up to 24 Gbps [\ref{Ref: VLC Speed 24Gbps}].

However, LiFi offers a relatively small coverage area with a single access point (AP), of usually 2-3 \rm{m} diameter. Also, LiFi is susceptible to channel blockage caused by opaque objects such as human bodies and furniture. Combining the high transmission speed of LiFi and the ubiquitous coverage of WiFi, the hybrid LiFi and WiFi network (HLWNet) is gaining research momentum recently. Such a network can greatly improve network capacity over stand-alone LiFi or WiFi networks [\ref{Ref: 7145863}]. Meanwhile, load balancing (LB) is identified as one of the key challenges in HLWNets [\ref{Ref: X. Wu Hybrid Network Survey}], due to the fact that WiFi APs have a larger coverage area but a lower system capacity than LiFi APs. When applying the signal strength strategy (SSS)\footnote{The SSS connects the user to the AP that provides the highest received signal power.} as if in a heterogeneous network, the users would be more likely to be served by WiFi than LiFi, making the WiFi system prone to overload. As a result, LB becomes essential and paramount to HLWNets, even though the users and their traffic demands are uniformly distributed in geography.

\subsection{Related Works}

\begin{table*}
\renewcommand{\arraystretch}{1.2}
\centering
\setlength\tabcolsep{1.5 mm}
\footnotesize
\caption{Comparison of Relevant Studies (MF: Max-Min Fairness; PF: Proportional Fairness; EPF: Enhanced PF).}
\begin{tabular}{|m{0.5cm}|m{1.2cm}|m{2.9cm}|c|c|c|c|m{4.2cm}|}
\hline
\textbf{Ref.} & \multicolumn{2}{l|}{\textbf{Method}} & \textbf{Optimality} & \textbf{Runtime} & \textbf{Fairness} & \textbf{LiFi AP density}\footnote{} & \textbf{Remark} \\
\hline
\multirow{2}{*}{[\ref{Ref: X. Li Cooperative LB TCOM}]} & \multicolumn{2}{l|}{Centralized optimization} & Optimal & Extremely High & \multirow{2}{*}{PF} & \multirow{2}{*}{16} & \multirow{2}{*}{Need to solve an NP-hard problem}  \\
\cline{2-5} & \multirow{5}{*}{\vtop{\hbox{\strut Iterative}\hbox{\strut method}}} & Distributed optimization & \multirow{5}{*}{Near-optim.} & \multirow{5}{*}{High} & &  &  \\ 
\cline{1-1} \cline{3-3} \cline{6-8}
[\ref{Ref: Y. Wang LB Game TWC}] & & Game theory & &  & MF, PF, EPF & 4, 16 & The number of iterations exponentially increases with the network size \\ 
\cline{1-1} \cline{3-3} \cline{6-8}
[\ref{Ref: College Admission Model TWC}] & & College admission model &  &  & $-$ & 1 & Solve a joint AP assignment and PA problem under constraints\\ 
\hline
[\ref{Ref: X. Wu AP Selection TCOM}] & \multirow{3}{*}{\vtop{\hbox{\strut Rule-based}\hbox{\strut decision} \hbox{\strut making}} }  & Fuzzy logic (FL) & Sub-optim. & \multirow{3}{*}{Medium} & PF & 16 & Lack adaptiveness due to fixed rules \\ 
\cline{1-1} \cline{3-4} \cline{6-8}
[\ref{Ref: Globecom Conference}]& & Mixed FL and optim. & Near-optim. & & PF & $-$ & Optimize resource allocation in the decision-making procedure of FL  \\ 
\hline
[\ref{Ref: RL for LB SJ}] & \multirow{3}{*}{\vtop{\hbox{\strut Learning}\hbox{\strut -aided} \hbox{\strut method}}} & Reinforcement learning & \multirow{3}{*}{Near-optim.} & \multirow{3}{*}{Low} & PF & 4  & Need retraining when the user number changes \\
\cline{1-1} \cline{3-3} \cline{6-8}
This work & & A-TCNN (DNN-aided) & & & PF & 4, 9 & Adapt to varying user numbers without the need for retraining \\
\hline
\end{tabular}\label{Table: Contributions}
\end{table*}

The conventional LB methods for HLWNets can be classified into three categories: centralized optimization [\ref{Ref: X. Li Cooperative LB TCOM}-\ref{Ref: Mobility-aware LB JOCN}], iterative methods [\ref{Ref: Y. Wang LB Game TWC}], [\ref{Ref: College Admission Model TWC}] and rule-based decision-making [\ref{Ref: Y. Wang Fuzzy Logic Handover ICC}-\ref{Ref: LB for Hybrid Networks}]. In [\ref{Ref: X. Li Cooperative LB TCOM}], a centralized optimization algorithm was proposed to achieve the proportional fairness (PF) between users. In [\ref{Ref: W. Ma Global Optimization LB CL}], the LB issue is formulated as mixed integer nonlinear programming (MINLP) problem considering users' data rate requirements and quality of service (QoS). In [\ref{Ref: Mobility-aware LB JOCN}], mobility-aware LB for both single transmission and multiple transmission modes was considered and then solved by the joint optimization method. Different from the above centralized optimization methods with their excessive computational complexity, autonomous optimization based on iterations can achieve near-optimal results, and to some extent reduce the complexity. In [\ref{Ref: Y. Wang LB Game TWC}], an iterative method based on evolutionary game theory (GT) was proposed for LB with multiple fairness functions considered. This iterative optimization method takes blockages, random orientation of LiFi receivers and users' data rate requirements into account. Taking power allocation (PA) into account, the authors in [\ref{Ref: College Admission Model TWC}] proposed college admission model based iterative method for joint AP assignment and power allocation problems in the three-tier hybrid VLC/RF network. However, the iterative method still requires a substantial amount of processing power. Rule-based decision-making method was proposed to further reduce the processing power [\ref{Ref: Survey on Fuzzy Logic}]. In [\ref{Ref: Y. Wang Fuzzy Logic Handover ICC}], the authors introduced a fuzzy logic (FL) based dynamic handover scheme for hybrid LiFi/RF network which considers user speed, data rate requirements, and time-varying signal-to-noise ratio (SNR) information, but without the optimization of throughput or user fairness. In [\ref{Ref: X. Wu AP Selection TCOM}], an FL based method was proposed to split the LB problem into two stages: i) determine the user which should be connected to WiFi; ii) allocate time resources for the remaining users in the stand-alone WiFi or LiFi network. Considering user mobility and light path blockage, the authors in [\ref{Ref: LB for Hybrid Networks}] proposed an FL based algorithm to reduce the complexity of the joint LB optimization problem in HLWNets. Although the rule-based decision-making method can handle the LB issue with a significant reduction of computational complexity, optimality is not achieved. Therefore, optimization based LB methods can provide optimal or near-optimal solutions, but centralized or iterative optimization algorithms require excessive computational complexity. In contrast, decision-making based methods, which demonstrate much lower complexity, compromise part of optimality.

\footnotetext[2]{The number of LiFi APs per coverage area of one WiFi AP.}

Unfortunately, the above conventional LB methods fail to achieve near-optimal performance and low computational
complexity at the same time, while the latter is crucial for meeting the zero-perceived latency requirement (0.1 ms) [\ref{Ref: 6G Vision}] in the sixth generation (6G) of wireless communication technology. For this reason, machine learning (ML) has recently enjoyed enormous attention to tackle the LB issue in heterogeneous networks [\ref{Ref: ML for Heterogeneous Survey}], and most of these approaches are based on reinforcement learning (RL). A few attempts have also been made to develop RL-aided LB methods for HLWNets [\ref{Ref: Q-learning for LB}-\ref{Ref: RL for LB SJ}]. In [\ref{Ref: Q-learning for LB}], a model-free and value-based Q-learning algorithm was proposed to maximize the network throughput. However, this method is infeasible to handle continuous state spaces, making it difficult to tackle densely deployed APs and users. In [\ref{Ref: DQN for LB}], Q-learning was combined with deep neural network (DNN) to jointly optimize bandwidth, power, and users association. Such a method is able to deal with discrete and continuous state spaces. In [\ref{Ref: RL for LB SJ}], a policy-based reinforcement learning (RL) algorithm was designed to determine an optimal AP assignment strategy, with the aim of maximizing the network throughput as well as the users' satisfaction and fairness. Compared with the value-based Q-learning methods, this policy-based method achieves a faster convergence. However, the above RL-aided methods all consider fixed numbers of users. When a new user joins the network or an existing user disconnects, these methods would need the retraining process to update the Q table or neural network, significantly limiting their practicability. Motivated by this, we aim to design an adaptive LB scheme for HLWNets which enables a state-of-the-art trade-off between optimality and complexity.

\subsection{Contributions}
In this paper, a novel DNN-aided LB method is proposed to tackle the LB issue in HLWNets, without the need for retraining when the number of users changes. The traditional DNN structure has fixed numbers of inputs and outputs, and thus is difficult to handle different numbers of users. To address this challenge, we develop a novel DNN structure named adaptive target-condition neural network (A-TCNN). The main contributions are explicitly contrasted to the existing literature in Table \ref{Table: Contributions}, and summarized as follows:


\begin{itemize}
\item A novel A-TCNN based LB approach is developed to tackle different numbers of users without the need for retraining. Specifically, the proposed A-TCNN introduces an adaptive mechanism to map a smaller user number to a larger one, followed by outputting the AP selection result for a target user based on the condition of other users.

\item Testing was carried out to validate the benefits of the proposed A-TCNN structure, which shows that compared with the method taken in test set, our A-TCNN method can achieve near equivalent performance in throughput and fairness. In addition, advantages of the proposed A-TCNN could be greater still when considering the effects of user clustering distributions.

\item A comprehensive comparison was conducted between the proposed A-TCNN and benchmarks, in terms of achievable throughput, fairness and computational complexity. Extensive results show that the proposed A-TCNN approach can achieve near-optimal network performance with a significantly reduced runtime up to 78 and 1400 times compared with two benchmarks.
\end{itemize}

The remainder of this paper is organized as follows. The system model is presented in Section II. The proposed method is elaborated in Section III. In Section IV dataset collection methods are summarized, and simulation results are shown in Section V. Finally, conclusion and future work are given in Section VI.

\textit{Notations:} Throughout the paper, we use lowercase italic letters to denote scalar variables, while bold lowercase and uppercase letters stand for vectors and matrices, respectively. The notations $\left|\cdot\right|$, $\left\| \cdot \right\|$ and $[\cdot]^{T}$ represent the absolute value, the Frobenius norm, and the transpose of a vector or matrix, respectively. $\mathbb{E}[\cdot]$ and $\mathrm{Var}[\cdot]$ are the expectation and variance values of a random variable. The operator $\lfloor \cdot \rfloor$ denotes the floor function, while ${\rm{mod}}(a, b)$ returns the remainder after a real number $a$ is divided by real number $b$. $\mathbb{R}^n$ represents the real coordinate space with a dimension of $n$. $\mathcal{N}(0, \sigma)$ denotes a Gaussian distribution with zero mean and standard variance $\sigma$.

\section{System Model}
The system model related to HLWNets is introduced in this section, including the network architecture, channel models and link capacity, and performance metrics.

\subsection{Network Architecture}
Fig. \ref{Fig:system model} shows an indoor HLWNet that consists of one WiFi AP and a number of LiFi APs. The WiFi AP is placed in the centre of the room on the ground, providing coverage to the entire room. The LiFi APs are arranged in a grid, with each AP embedded into a ceiling lamp, covering a confined area. Frequency reuse with a reuse factor of 4 is adopted to avoid inter-cell interference (ICI) among adjacent LiFi APs [\ref{Ref: Parallel LiFi}]. The ICI between further APs is trivial and can be neglected. The users are randomly located on the ground with a uniform distribution. The data rate requirements of the users are considered to be independent and identically distributed random variables. Without loss of generality, it is assumed that these variables follow a Gamma distribution with shape parameter $k$ and scale parameter $\theta$ [\ref{Ref: Gamma Distribution}]. Let $R_{j}$ denote the data rate required by the $j$-th user. The expected value of the average data rate requirement is $\bar{R} = k\theta$. Each user is connected to one AP, either WiFi or LiFi, while each AP can serve multiple users via time-division multiple access (TDMA).

\subsection{Channel Models and Link Capacity}
\begin{figure}[t]
\centering
\includegraphics[height=6cm,width=9cm]{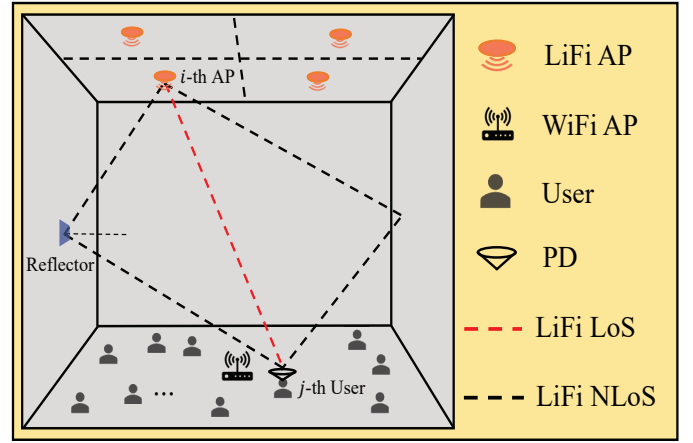}
\caption{Schematic diagram of an indoor HLWNets.}
\label{Fig:system model}
\end{figure}

Quasi-static channels are considered here. For WiFi, the log-distance path loss model in [\ref{Ref: WiFi Model}] is adopted. As for LiFi, the channel consists of the line-of-sight (LoS) and first-order non-line-of-sight (NLoS) paths, as illustrated in Fig. \ref{Fig:system model}. The corresponding expressions can be found in [\ref{Ref: LoS and NLoS} eq.(10) and eq.(12)]. The capacity of WiFi is bounded by Shannon capacity. With respect to LiFi, a tighter bound can be found in [\ref{Ref: Tight Bound JLT}] due to the non-negative real signals in LiFi. Let $\gamma_{i,j}$ denote the signal-to-noise ratio (SNR) of the link between AP $i$ and user $j$. The link capacity $C_{i,j}$ can be expressed as:
\begin{equation}
C_{i,j} = \left\{
\begin{aligned}
&{{B_i}\over{2}}{\log_2}\big(1 + {{e}\over{2\pi}} \gamma_{i,j}\big), \:\:\: \rm{for \: LiFi} \\
&{B_i}{\log_2}(1 + \gamma_{i,j}) , \quad\quad \rm{for \: WiFi}  \\
\end{aligned}
\right.
\end{equation}
where $e$ is the Euler’s number and $B_{i}$ denotes the bandwidth of AP $i$. 

\subsection{Performance Metrics}
Let $\Gamma$ denotes the achievable throughput, which is given by:
\begin{equation}
\Gamma = \sum\limits_{i \in \mathbb{S}}\sum\limits_{j \in \mathbb{U}}{\rho_{i,j}}{\chi_{i,j}}{C_{i,j}},
\end{equation}
where $\mathbb{S} = { \{i|i\in [1, 2, ..., N_{\rm{t}}] \} }$ and $\mathbb{U} = { \{j|j\in[1, 2, ..., N_{\rm{u}}] \} }$ denote the sets of APs and users, while $N_{\rm{t}}$ and $N_{\rm{u}}$ are the total numbers of APs and users, respectively; $\chi_{i,j} = 1$ indicates there is a connection between AP $i$ and user $j$, and otherwise $\chi_{i,j}=0$; $\rho_{i,j} \in [0, 1]$ denotes the portion of time resource that is allocated by AP $i$ to user $j$.

Let $S_j$ denote the satisfaction degree of user $j$, which can be expressed as:
\begin{equation}
{S_{j}} = \min \bigg\{ \frac{ {\sum_{i \in \mathbb{S}}{{\rho_{i,j}}{\chi_{i,j}}{C_{i,j}}}} }{{R_j}}, 1\bigg\}.
\end{equation}

The fairness among users, denoted by $\xi$, is commonly measured by Jain's fairness index [\ref{Ref: Jain's Fairness}]:
\begin{equation}
\xi =\frac{\left(\sum_{j \in \mathbb{U}}S_j\right)^2}{N_{\rm{u}}\sum_{j \in \mathbb{U}}S_j^2}.
\end{equation}

\section{Proposed DNN-Aided Framework for LB}

\begin{figure*}[htb]
\centering
\includegraphics[width=18cm]{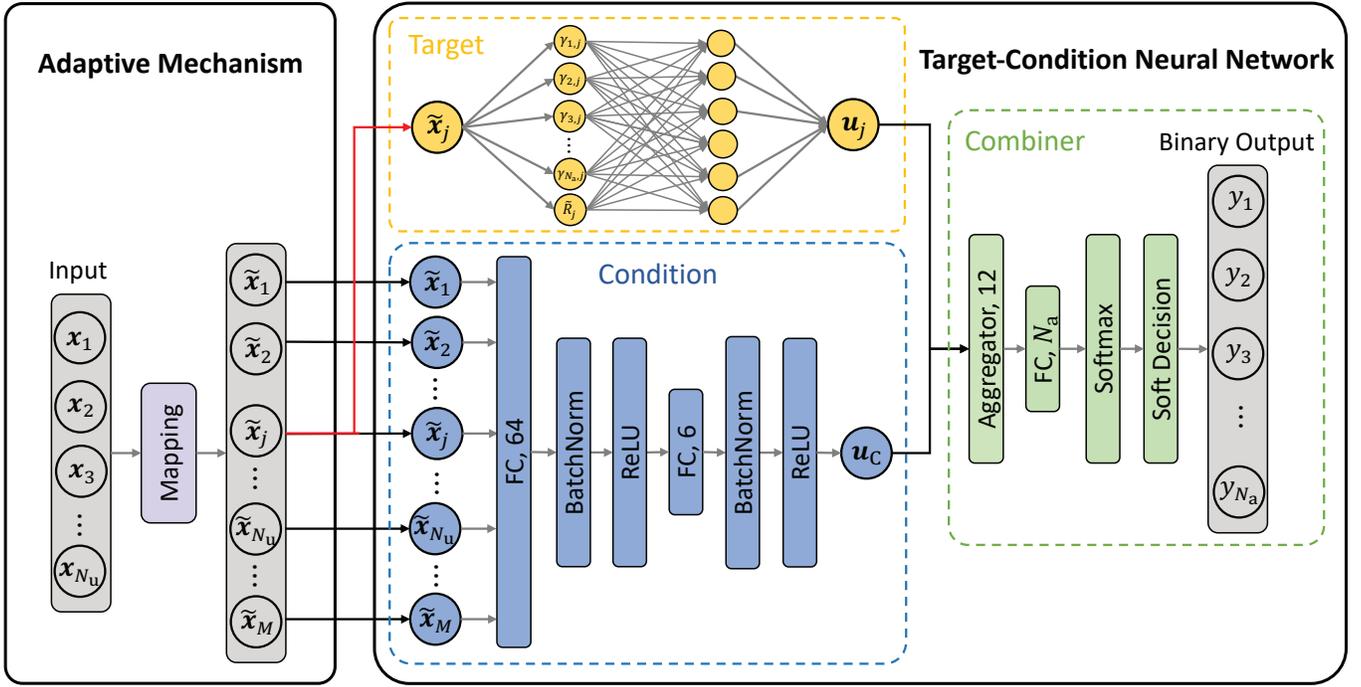}
\caption{Block diagram of the proposed A-TCNN.}
\label{Fig:Adaptive TCNN}
\end{figure*}

In this section, we present a novel DNN-aided LB method for HLWNets, which consists of two key components: the target-condition neural network (TCNN) and adaptive mechanism with respect to input user number. As illustrated in Fig. \ref{Fig:Adaptive TCNN}, the block diagram of the proposed A-TCNN framework includes two solid rounded rectangles. The right block represents the TCNN model, and the left one refers to the adaptive mechanism with respect to number of working users in the HLWNets. A detailed explanation on the proposed A-TCNN model for LB is given in subsections III-A and III-B. Finally, we introduce our training and evaluation procedure for the A-TCNN model. 

\subsection{Target-Condition Neural Network}
For the conventional DNN model, a basic solution is to input the collected data including SNRs and data rate requirements for all users, and then to simultaneously estimate their optimal AP assignments, which is referred as DNN in this work. However, it does not support the adaptation to the number of users. When applying the adaptive mechanism in DNN model, the mirroring users split from the same user may result in different output results, which is conflicted in nature. Without an adaptive mechanism, the trained DNN network only working on a specific user case lacks of scalability. To solve this limitation, we focus our attention on a single user which needs to be estimated in the output and set all other users as condition, shown in the inner dashed rounded rectangle of Fig. \ref{Fig:Adaptive TCNN}. In our implementation, the TCNN is separated into three parts, namely \textit{target}, \textit{condition} and \textit{combiner}, respectively.

\subsubsection{Target}
For the target block, the focus is on user $j$. User $j$ contains two kinds of data: SNRs between candidate AP $i$ and user $j$ pairs, and the partial data rate requirement of user $j$ after the mapping. Therefore, the fed vector of target is expressed as $\boldsymbol{\tilde{x}}_j = [\gamma_{1,j}, \gamma_{2,j}, ..., \gamma_{N_{\rm{a}},j}, \tilde{R}_j]^T \in \mathbb{R}^{N_{\rm{a}}+1}$. Then, a fully-connection (FC) layer with 6 neurons is employed to process $\boldsymbol{\tilde{x}}_j$, which is illustrated in the rounded rectangle with yellow dash in Fig. \ref{Fig:Adaptive TCNN}. Let us denote the weights and bias of FC layer by $\mathbf{W}_1 \in \mathbb{R}^{(N_{\rm{a}}+1)\times 6}$ and $\boldsymbol{b}_1 \in \mathbb{R}^6$. Therefore, the output of target part can be expressed as:
\begin{equation}
\boldsymbol{u}_j = {\mathbf{W}_1}{\boldsymbol{\tilde{x}}_j} + \boldsymbol{b}_1.  
\end{equation}

\subsubsection{Condition}
In the condition block, we deem all users as condition to input into the TCNN, which is an arrangement of entire SNR and data rate requirement parameters (including the focused user $j$ mentioned in the above target block). Let us denote the input of condition as $\boldsymbol{\bar{x}}_{\rm{C}}$, given as $\boldsymbol{\bar{x}}_{\rm{C}} = [{\boldsymbol{\tilde{x}}^T_1}, {\boldsymbol{\tilde{x}}^T_2}, ..., {\boldsymbol{\tilde{x}}^T_M}]^T \in \mathbb{R}^{M(N_{\rm{a}}+1)}$, where $M$ is a default maximum user number of input. For example, the maximum number of working users that can be estimated by the proposed TCNN is 50 if we fix $M$ as 50. To reduce the dimension of the condition while extracting the principal components, two FC layers with 64 and 6 neurons are utilized here, and the corresponding network parameters are denoted as $\mathbf{W}_2 \in \mathbb{R}^{64 \times M({N_{\rm{a}}}+1)}$, $\boldsymbol{b}_2 \in \mathbb{R}^{64}$ and $\mathbf{W}_3 \in \mathbb{R}^{6 \times 64}$, $\boldsymbol{b}_3 \in \mathbb{R}^6$, respectively. After that, the rectified linear unit (ReLU) is used as the activation function, defined as ${f_{\rm{ReLU}}}(\boldsymbol{x}) = {\rm{max}}({\rm 0}, \boldsymbol{x})$. Note that the eigenvalues of the FC layer's output would gradually approach the saturation interval of the activation function, causing gradients to tend to vanish during the training process. Therefore, the output from FC must to be normalized using a batch normalization (BN) layer to enable the eigenvalues distribution to stay within the standard normal distribution [\ref{Ref: Batchnorm}], which is more sensitive to the activation function. With this step, vanishing gradients can be avoided and training convergence rate improved. Here, the BN operation can be formulated as follows:

\begin{equation}
f_{\mathrm{BN}}(\boldsymbol{x}) = \frac{ \boldsymbol{x} - \mathbb{E}\left [ \boldsymbol{x} \right ]} {\sqrt{\mathrm{Var}\left [ \boldsymbol{x} \right ] + \epsilon }} \mu + \nu
\end{equation}
where the $\epsilon$ is a small constant close to zero to avoid division by zero, $\mu$ and $\nu$ are the trainable factors for scaling and shifting the distribution, which are set as 1 and 0 initially. Finally, the output of the condition block is as follows:
\begin{equation}
\small
\boldsymbol{u}_{\rm{C}} = {f_{\rm{ReLU}}}\left\{ {{f_{\rm{BN}}}\left[{{\mathbf{W_3}}\left\{ {{f_{\rm{ReLU}}}\left[ {{f_{{\rm{BN}}}}\left( {{\mathbf{W_2}}{{\bar{\boldsymbol{x}}}_{\rm{C}}} + {\boldsymbol{b_2}}} \right)} \right]} \right\} + \boldsymbol{b_3}} \right]} \right\}
\end{equation}

\subsubsection{Combiner}
In the combiner part, the subtracted features from the target and condition are concatenated (Cat) together in the aggregator and fed into the next layers. Here, the operation of Cat is expressed as ${\mathrm{Cat}}({\boldsymbol{u}_j}, {\boldsymbol{u}_{\rm{C}}}) = [{\boldsymbol{u}^T_j}, {\boldsymbol{u}^T_{\rm{C}}}]^T \in \mathbb{R}^{12}$. Similarly, an FC layer with weights matrix $\mathbf{W}_4 \in \mathbb{R}^{N_{\rm{a}} \times 12}$ and bias vector $\boldsymbol{b}_4 \in \mathbb{R}^{N_{\rm{a}}}$ is introduced here, followed by the probability-based activation function softmax. The softmax function is defined as ${f_{\rm{softmax}}}(\boldsymbol{x}) = e^{\boldsymbol{x}} / {\sum}e^{\boldsymbol{x}}$. Lastly, the focused user $j$ connects the corresponding AP $i^{\ast}$ with the largest possibility using a soft decision maker. The estimated optimal AP $i^{\ast}$ in the combiner can be calculated as:
\begin{equation}
{i^*} = \arg \mathop {\max}\limits_{i \in \mathbb{S}} \left\{ {{f_{{\rm{softmax}}}}\left[ {{{\mathbf{W}}_4}\left( {\rm{Cat}}({\boldsymbol{u}_j}, {\boldsymbol{u}_{\rm{C}}} \right) + {\boldsymbol{b}_4}} \right]} \right\}, 
\end{equation}
and the final estimated output is ${\boldsymbol{\hat{y}}_j} = [y_1, y_2, ..., {y_{N_{\rm{a}}}}]$, which satisfies that ${\boldsymbol{\hat{y}}}_j(i) = 1$ when $i = i^*$, otherwise ${\boldsymbol{\hat{y}}_j}(i) = 0$ for $\forall i \in \mathbb{S}$ and $\forall j \in \mathbb{U}$. Note, the estimated binary output ${\boldsymbol{\hat{y}}_j}(i)$ indicates the AP selection choice for each user $j$ with respect to AP $i$ in the HLWNets, where ${\boldsymbol{\hat{y}}_j}(i) = 1$ means there should be a connection, otherwise no connection.

\subsection{Adaptive Mechanism}
In the proposed TCNN, the estimation target is focused from the entire users to a single user, which is estimated in the combiner. However, the dimension of input fed into the condition block is unchangeable, i.e., the proposed TCNN can only work on $M$ users number case. To provide an adaptive functionality with respect to the estimated user number, we further present an adaptive mechanism in this subsection, which is illustrated in the left solid rounded rectangle in Fig. \ref{Fig:Adaptive TCNN}. 

In general, the proposed adaptive mechanism can transform $N_{\rm{u}}$ users into $M$ mapped users. First, let us denote the user-centric data (including considerable SNRs $\gamma_{i,j}$ and data rate requirement $R_j$ for user $j$) as $\boldsymbol{x}_j = [\gamma_{1,j}, \gamma_{2,j}, ..., \gamma_{N_{\rm{a}},j}, R_j]^T \in \mathbb{R}^{N_{\rm{a}}+1}$. Then, a set of $\boldsymbol{x}_j$ are sent into the mapping operator, in which the mirroring users that equally share the data rate requirement $R_j$ while remaining the same SNR values are created. Here, we use $\boldsymbol{\tilde{x}}_j \in \mathbb{R}^{N_{\rm{a}}+1}$ indicates the mirroring user $j$, where $j$ is an integer ranging from 1 to $M$. For example, the user number ${{N_{\rm{u}}}}$ is 20 here and $M$ is set as 30. After mapping, the mirroring data matrix can be expressed as $\left [\boldsymbol{\tilde{x}}_1, \boldsymbol{\tilde{x}}_2, ..., \boldsymbol{\tilde{x}}_{20}, \boldsymbol{\tilde{x}}_{21}, ..., \boldsymbol{\tilde{x}}_{30} \right ]$, here $\boldsymbol{\tilde{x}}_j = [\gamma_{1,j}, \gamma_{2,j}, ..., \gamma_{N_{\rm{a}},j}, R_j/2]^T$ for $\forall j \in [1, 2, ..., 10]$ and $[21, 22, ..., 30]$, and $\boldsymbol{\tilde{x}}_j = [\gamma_{1,j}, \gamma_{2,j}, ..., \gamma_{N_{\rm{a}},j}, R_j]^T$ for $\forall j \in [11, 12, ..., 20]$. A general algorithm of the adaptive mechanism is shown in Alg. \ref{Alg:Adaptive_Mechanism}. Finally, an operation is conducted to separate the target and condition, and the corresponding streams are denoted as $\boldsymbol{\tilde{x}}_j$ and $\boldsymbol{\bar{x}}_{\rm{C}} = [{\boldsymbol{\tilde{x}}^T_1}, {\boldsymbol{\tilde{x}}^T_2}, ..., {\boldsymbol{\tilde{x}}^T_M}]^{T}$, respectively. 

\begin{algorithm}[t]\label{Alg:Adaptive_Mechanism}
\SetAlgoLined
\KwIn{${\boldsymbol{x}_1}, {\boldsymbol{x}_2}, ..., {\boldsymbol{x}_{N_{\rm{u}}}}, {N_{\rm{u}}}, M$}
\KwOut{$\boldsymbol{\tilde{x}}_1, \boldsymbol{\tilde{x}}_2, ..., \boldsymbol{\tilde{x}}_M$}
Calculate $a =\lfloor{M/{{N_{\rm{u}}}}} \rfloor $, and $b = {\rm{mod}} (M, {{N_{\rm{u}}}}) $ \;
\For {$j = 1$ {\rm{to}} $M$}{ 
\eIf{$ j \leqslant b$}{
$\boldsymbol{\tilde{x}}_j \Leftarrow [\gamma_{1,j}, \gamma_{2,j}, ..., \gamma_{N_a+1,j}, R_j/(a+1)]$\;
}
{\eIf{$ j \leqslant {N_{\rm{u}}}$}
{$\boldsymbol{\tilde{x}}_j \Leftarrow [\gamma_{1,j}, \gamma_{2,j}, ..., \gamma_{N_a+1,j}, R_j/a]$\;}
{$j^\ast = {\rm{mod}}(j-1, N_{\rm{u}}) + 1 $ \;
$\boldsymbol{\tilde{x}}_j \Leftarrow [\gamma_{1,j^\ast}, \gamma_{2,j^\ast}, ..., \gamma_{N_a+1,j^\ast}, R_j/a]$\;}
}}
\caption{Adaptive Mechanism.}
\end{algorithm}

\subsection{Training and Testing}
In our training phase, one key feature is that adaptive user number can be chosen to train the network. Assuming that the user number $N_{\rm{u}}$ ranges from 5 to $M$ and dataset collection can be resolved in the next section, our training process can be introduced by two aspects.

\subsubsection{Fixed Numbers of Users} For the case of TCNN with specific user number of $N_{\rm{u}}$, we first collect $N$ batches for training and the batch size is $T$. In one batch, each column denotes a sample showing as $\boldsymbol{\bar{x}}_{\rm{C}} = [{\boldsymbol{\tilde{x}}^T_1}, {\boldsymbol{\tilde{x}}^T_2}, ..., {\boldsymbol{\tilde{x}}^T_M}]^{T}$. Note that the fed data $\boldsymbol{\bar{x}}_{\rm{C}}$ containing $\gamma_{i,j}$ and $R_j$ needed to be normalized separately, since $\gamma_{i,j}$ and $R_j$ are drawn from different distributions. Here, linear normalization is adopted for $\gamma_{i,j}$ while logarithmic normalization is utilized for $R_j$.\footnote{For SNRs in dB scale, most values follow a normal distribution, while $R_j$ with a gamma distribution is mainly centred around the mean value of $k\theta$. Providing that the minimum and maximum values for $R_j$ are 1 and 1000 Mbps respectively, and the mean value is 100 Mbps, linear normalization for $R_j$ would compress most $R_j$ to a small range, which would slow down learning. As a result, we adopted the logarithmic normalization, which can better scale most $R_j$ in range of [0, 1].} In addition, all batches are normalized with the same maximum and minimum values, which are obtained from the training dataset. After that, normalized $N$ batches are grouped together and fed into the TCNN. The ground truth labels can be denoted as $\boldsymbol{y}_j$, which is the correct AP assignment results for the input user $j$ and has been saved in the output of dataset. During training process, we use stochastic learning rather than batch learning as it can speed up learning, particularly on large redundant datasets [\ref{Ref: Efficient BackProp}]. We selected the mean square root (MSE) loss function to maximize the agreement between the estimation from the proposed TCNN and the label values, formulated as: 
\begin{equation}\label{Equation:MSE Loss}
L_{\mathrm{MSE}}\left ( \theta  \right )=\frac{1}{T}\sum_{j=1}^{T}\left\| {\boldsymbol{y}_j}-{\hat{\boldsymbol{y}}_j}\right\|^{2},
\end{equation}
where the $\theta$ is the trainable parameter. During training process, network parameter $\theta$ is updated and optimized using adaptive moment estimation (Adam) scheme [\ref{Ref: Adam}] for each batch over $T$ batch samples. As a result, the updating of $\theta$ can be formulated as follows:
\begin{equation}
\theta \Leftarrow \theta -\eta \nabla L_{\mathrm{MSE}}\left ( \theta  \right ),
\end{equation}
where $\eta$ is the learning rate during training process, and $\nabla L_{\mathrm{MSE}}(\theta)$ means the gradient of the loss function with respect to $\theta$. 

\subsubsection{Adaptive Numbers of Users}
To adapt to a variable number of users, we aim to convert the dataset consisting of $N_{\rm{u}}$ users into the mapped dataset with $M$ users so that the TCNN model with $M$ input users is still working for a smaller user count. The detailed illustration of adaptive mechanism can be referred to Alg. \ref{Alg:Adaptive_Mechanism}. 

The testing phase includes loss testing and performance evaluation. As for the loss testing, 256 samples are collected for each user example. The computation process of testing is the same as the training, while the loss will not be back-propagated to update parameters of the neural network. The model can be verified if it overfits the training set through the MSE loss testing formulated in (\ref{Equation:MSE Loss}). In the evaluation stage, two network metrics are adopted in this work: \textit{accuracy} and \textit{performance gap}. 
\begin{itemize}
\item \textit{Accuracy:} We define accuracy as the ratio between the number of users with correct AP connection estimated by A-TCNN model and the total number of tested users.
Note correct AP connection is evaluated based on the ground truth of dataset. Besides, users at different positions are sequentially fed into the TCNN model. 
\item \textit{Performance gap:} In HLWNets, AP assignment results have a significant influence on the achievable throughput of hybrid network. Therefore, we define performance gap as the achievable network throughput gap between the estimated AP selection and the real AP assignment results, where the same optimal PA is achieved for comparison.
\end{itemize}

\section{Dataset Collection}
In this section, we introduce and discuss a number of potential dataset collection methods for the proposed A-TCNN. These dataset collection methods are four-fold: \textit{global optimization}, \textit{direct decision-making}, \textit{iteration based}, and \textit{mixed decision-making and optimization} methods.

\subsection{Global Optimization Method}
The aim of global optimization is to maximize the user satisfaction with proportional fairness [\ref{Ref: PF}].
Global optimization is an optimal solving algorithm which provides the upper bound for the problem formulation, thus it is also termed as exhaustive search or brute force search. The problem formulation of global optimization for LB in HLWNets can be expressed as:

\begin{equation}\label{eq:Problem}
\begin{array}{l}
\!\!\!\!{\max\limits_{\chi_{i,j},\rho _{i,j}}\quad\! \sum\limits_{j \in \mathbb{U}}{\log S_{j}}}\\
{{\rm{s}}{\rm{.t}}{\rm{.}}\quad\quad\! \sum\limits_{i \in \mathbb{S}} {\chi_{i,j}} = 1} ,\forall j \in \mathbb{U};\\
\quad\quad\quad \; \sum\limits_{j\in\mathbb{U}} {\chi_{i,j}{\rho_{i,j}} \le 1} ,\forall i \in \mathbb{S};\\
\quad\quad \quad\; 0 \le {\rho _{i,j}} \le 1,\forall\; i,j;\\
\quad\quad \quad\; {\chi_{i,j}} \in \{ 0,1\} ,\forall\; i,j,
\end{array}
\end{equation}

Here (\ref{eq:Problem}) is similar to problem formulation given in [\ref{Ref: X. Li Cooperative LB TCOM}], but the objective function is revised with consideration of users' satisfaction. As shown in (\ref{eq:Problem}), assigning AP selections is an NP-hard problem [\ref{Ref: NP-hard Problem}] and PA in a stand-alone network is a nonlinear programming problem. Although it has been shown that MINLP can be solved by the OPTI toolbox, nonlinear branch-and-bound algorithm based global optimization solvers (e.g. BONMIN, SCIP) take a long time to return global results [\ref{Ref: MINLP}]. Particularly in the case of dense deployment of APs and users, the computational complexity is exponentially increasing with regard to the network size, which is far from a wise deployment for a real-time and user-centric network.

\subsection{Direct Decision-Making Method} 
For direct decision-making method, the object is to make an artificial but reasonable rule set according to the existing social or natural system. In [\ref{Ref: X. Wu AP Selection TCOM}], the authors presented an FL based LB method for HLWNets, where the ICI is solved by introducing SNR variance and activity of adjacent AP to define how likely the user is affected by the ICI. However, in this paper, we consider frequency reuse to avoid ICI [\ref{Ref: Parallel LiFi}]. As a result, we revise the original FL rules for a fair comparison with benchmarks and the new fuzzy rule is listed in Table \ref{Table: Fuzzy Rule}. Note in the FL method, we adopt SSS as the AP assignment method in the second stage of LB, thus only the nearest LiFi AP providing biggest SNR will be considered for each user in the fuzzy rule. Though this kind of heuristic decision-making method can significantly reduce the processing time, it is unknown how far the solution is from achieving the optimal optimization results. In addition, the designed rules cannot adapt to changes in the environment such as changing user number and/or network size, as is seen in HLWNets.

\begin{table}
\renewcommand{\arraystretch}{1.2}
\centering
\setlength\tabcolsep{1.6 mm}
\caption{Fuzzy Rules for the FL Method (Req.: Required Data Rate; Ava.: Availability of WiFi or LiFi AP Resource; Sel.: Selection)}
\scriptsize
\begin{tabular}{|c|c|c|c|c|c|c|} 
\hline
\textbf{Rule} & \textbf{Req.} & \textbf{WiFi SNR} & \textbf{LiFi SNR} & \textbf{WiFi Ava.} & \textbf{LiFi Ava.} & \textbf{WiFi Sel.} \\ 
\hline
 1 & Low & $-$ & $-$ & \textit{not} Low & $-$ & High \\ 
\hline
 2 & \textit{not} Low & $-$ & $-$ & \textit{not} Low & $-$ & Medium  \\ 
\hline
 3 & $-$ & $-$ & $-$ & Low & \textit{not} Low &  Low \\ 
\hline
 4 & $-$ & Low & \textit{not} Low & Low & Low & Low  \\ 
\hline
 5 & $-$ & Low & Low & Low & Low & Medium  \\ 
\hline
 6 & $-$ & \textit{not} Low & Low & Low & Low & High  \\ 
\hline
 7 & $-$ & \textit{not} Low & \textit{not} Low & Low & Low & Medium  \\ 
\hline
 8 & High & $-$ & High & $-$ & $-$ & Low  \\ 
\hline
 9 & \textit{not} High & High & $-$ & \textit{not} Low & $-$ & High \\
\hline
\end{tabular}\label{Table: Fuzzy Rule}
\end{table}

\subsection{Iterative Method} 
Another typical strategy which is capable of solving LB problem utilizes a number of iterations to reach a steady state, which means that a near-optimal result is achieved. Here, we select GT based LB method [\ref{Ref: Y. Wang LB Game TWC}] as iteration based baseline to compare. During the GT process, each player occurs randomly and follows the rule that the player with lower value of payoff would be more likely to change its strategy, which is termed as `mutation and selection mechanism' rule [\ref{Ref: Y. Wang LB Game TWC}]. The Nash equilibrium is obtained when all players can not continue mutation and selection for one loop. However, it is not always optimal as no single player can improve its payoff does not guarantee that the entire players can further globally maximize their payoffs. Besides, this GT approach is highly sensitive to initial states. 

\subsection{Mixed Decision-Making and Optimization Method}
In this paper, we extend our previous work [\ref{Ref: Globecom Conference}] to LB for HLWNets and introduce a revised FL based method which combines decision-making and optimization to provide a better trade-off between optimality and complexity. This method is termed as FL-OPT in this paper. The proposed mixed method for dataset collection considers user satisfaction maximization in each decision step. In general, the mixed LB method follows two principles: i) for users candidates, each user $j$ first chooses the AP providing the biggest $S_{j}$ as candidates; ii) for chosen AP candidates, connect a pair of AP and user with the highest output score yielded in fuzzy logic system. Therefore, the proposed dataset collection method has inherent low-complexity advantage of decision-making and performance enhancement characteristic of optimization method. Specifically, the satisfaction optimization problem providing a specific set of $\chi_{i,j}$ can be formulated as:
\begin{equation}\label{eq:Problem2}
\begin{array}{l}
\max\limits_{\rho_{i,j}} \;\quad\sum\limits_{j \in \mathbb{U}} {\log S_{j}}\\
\;\;{\rm{s}}{\rm{.t}}{\rm{. }}\quad\;\sum\limits_{j\in\mathbb{U}} {\chi_{i,j}{\rho_{i,j}} \le 1} ,\forall i \in \mathbb{S};\\
\quad\quad \quad\;\; 0 \le {\rho _{i,j}} \le 1,\forall\; i,j,\\
\end{array}
\end{equation}
By creating the Lagrangian function and using the Karush–Kuhn–Tucker (KKT) conditions [\ref{Ref: Karush}] and [\ref{Ref: Kuhn-Tucker}], the optimal PA results of $\rho_{i,j}$ for a given case of $\chi_{i,j}$ can be calculated as:
\begin{equation}\label{Equation:PA results}
{{\rho_{i,j}} \simeq { 1 \over {\sum\limits_{j\in\mathbb{U}}{\chi_{i,j}}}}}
,\quad \forall \; i.
\end{equation}
where $\simeq$ denotes the optimal PA results can be approximately equal to the right term in (\ref{Equation:PA results}). The proof of (\ref{Equation:PA results}) is provided in Appendix A. In the decision making process, the approximate PA result shown in (\ref{Equation:PA results}) is adopted. The fuzzified boundaries are set as (10, 70) dB and (0, 10000) Mbps, also fuzzified breakpoints are (25, 40, 44) dB and (0, $\bar{R}$, $2\bar{R}$) Mbps for SNR and data rate requirements respectively. For the sake of simplicity, the detailed technical descriptions and other parameter values are referred to our previous work [\ref{Ref: Globecom Conference}].

In this paper, choosing the global optimization method as a benchmark is infeasible due to the excessive computations. For the direct decision-making method, its optimality is compromised. The complexity and runtime of mixed decision-making and optimization method is lower than the iteration based method, thus we adopt the mixed decision-making and optimization method to generate the dataset for training and testing the A-TCNN, which is analysed in the next section.

\section{Simulation Results}

\subsection{Simulation Setup}
In this section, the Monte Carlo simulations are conducted to evaluate the proposed A-TCNN structure versus its benchmarks: SSS, GT [\ref{Ref: Y. Wang LB Game TWC}], and FL [\ref{Ref: X. Wu AP Selection TCOM}] methods, in which the network throughput and users' Jain's fairness are chosen as wireless network metrics for comparison. Here we consider two general square room sizes with side length of 5 m and 9 m, where LiFi AP number is 4 and 9, respectively. For each LiFi, the transmitted optical power is 3 W, and breakpoint distance for WiFi is 3 m. Simulation setup is coded on MATLAB R2021a. Other parameters can be referred to [\ref{Ref: Globecom Conference}].

\begin{figure}[t]
\centering
\includegraphics[width=9cm]{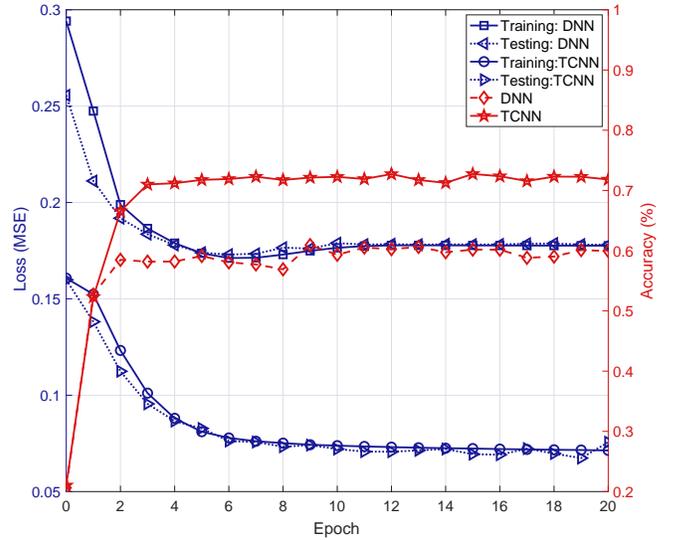}
\caption{Training loss and estimation accuracy for the proposed TCNN versus benchmark of DNN, where $N_{\rm{u}}$ = 50.}
\label{Fig:CDNN_versus_Global}
\end{figure}

\begin{figure}[t]
\centering
\includegraphics[height=7.5cm,width=9cm]{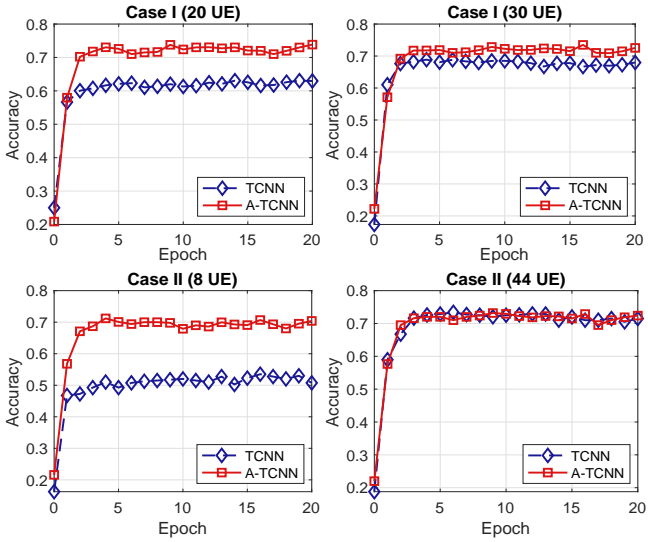}
\caption{Comparisons of accuracy for the proposed A-TCNN versus TCNN under various user numbers, where Case I picks two chosen user numbers (20 and 30) which has been used in the training process, while Case II adopts two unchosen user numbers (8 and 44).}
\label{Fig:Adaptive_versus_CDNN}
\end{figure}

The architecture of the A-TCNN has been shown in Fig. \ref{Fig:Adaptive TCNN}, where the main features of network (e.g. layer number, neutrons number and activation function) has been presented. For simplicity, here we consider 4 LiFi AP case for training discussion and presentation in Fig. \ref{Fig:CDNN_versus_Global} and Fig. \ref{Fig:Adaptive_versus_CDNN}. Also, $M$ is fixed as 50 for simplicity in this paper. During the offline training process, we set batch number $N$ as 1000 and batch size $T$ as 256. Note we pick user number as 5, 10, 15, ..., 50 (10 training cases equally share 1000 batch) for training. Value of momentum in Adam optimizer is 0.95 and learning rate is chosen as 0.0001. During the offline testing, batch size for validation is set as 256 to evaluate the testing loss and performance metrics. The training, testing and evaluation are coded in Python3, which is running in a PC with Linux operation system, Intel Core i7-10875H processor and NVIDIA 2060 GPU. Our training code is open-sourced at [\ref{Ref: Github}].

\subsection{Learning Network Metrics}
In this subsection, we first to evaluate the training process between TCNN and DNN. Note DNN is referred to the structure of the conventional DNN. As can be shown in Fig. \ref{Fig:CDNN_versus_Global}, the training loss of TCNN and DNN decrease with the increase of epoch, and finally converge. Here, each epoch takes about 8.3 s runtime for training TCNN without evaluating performance metrics. Note, training time of each epoch for A-TCNN is the same as TCNN as they have the same training structure in nature.
Totally, runtime of convergence is about 83 s for the proposed (A-)TCNN model, which is similar to DNN benchmark. Besides, it shows that both networks can constantly learn the relationship between input and output well because the same dataset can be seen with epoch index increases. For online testing, the testing loss approaches training loss curve well, which illustrates no over-fitting occurs during training. In other words, the trained networks both have a good generalization ability over unseen data. As for the training accuracy, the TCNN outperform DNN with about 12\% improvement when training is finished. The advantage of accuracy is explained that the increased dimension of input and output vector space in DNN challenges estimation mission.

Next, another experiment is conducted for the A-TCNN and pure TCNN with different user number structure in Fig. \ref{Fig:Adaptive_versus_CDNN}. Specifically, two cases are picked here, where Case I represents the chosen user number (20 and 30) belongs to the trained user number (e.g. 5, 10, ..., 50), and Case II represents the chosen user number (8 and 44) are not trained. In this figure, three significant clues can be found:

\begin{figure}[t]
\centering
\includegraphics[height=7.5cm, width=9cm]{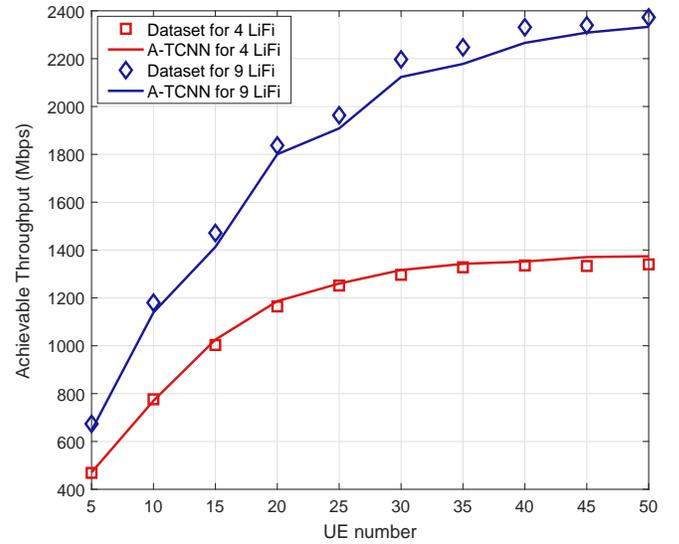}
\caption{Achievable throughput for the proposed A-TCNN model compared with training dataset.}
\label{Fig:Adaptive_Gap}
\end{figure}

\begin{itemize}
\item Denote the different user number in TCNN as $M_{\rm{C}}$. We first conclude that with the increase of $M_{\rm{C}}$ from 8 to 44, the accuracy of TCNN rises from 50\% to 70\%. The higher accuracy is mainly because the condition part with the increased dimension in TCNN can be extracted more features for learning, which inspires us that the structure of TCNN (i.e. maximum user number of $M_{\rm{C}}$) plays a dominant role in accuracy wise. In addition, previous Fig. \ref{Fig:CDNN_versus_Global} shows that TCNN with 50 user number structure can reach about 72\% accuracy. Compared with 70\% when $M_{\rm{C}}$ = 44, it is also indicated that the accuracy improvement is not significant when user number is bigger than 50. This drive us to choose $M$ as 50 for the proposed A-TCNN.

\item Second, accuracy of A-TCNN remains stable at level of 70\% for different user numbers. This means that the different user numbers of A-TCNN can reach a near accuracy level as TCNN with 50 UE number since the maximum default UE number $M$ in A-TCNN is fixed as 50. 

\item Last, another observation can be seen from comparison between Case I and Case II. For A-TCNN, both Case I (20 and 30 UE cases) and Case II (8 and 44 cases) have the same accuracy performance, which illustrates that the trained A-TCNN can not only estimate the seen user cases (e.g. 20 and 30), but also generates a good generalization ability over the unseen user cases. This indicates us no need to collect entire users' cases for training.
\end{itemize}

From the above observations, we show that the A-TCNN has a better accuracy and scalability. From communication network wise, we next evaluate its performance gap compared with the provided standard dataset. As presented in Fig. \ref{Fig:Adaptive_Gap}, we plot achievable throughput for the A-TCNN method using estimated AP assignments scheme versus dataset for different user number, where we evaluate performance gap in different HLWNets size. In general, the estimated curve of A-TCNN fits the standard points given in dataset well without big performance gap. It is worthy noting that smaller network size of 4 LiFi performs better than bigger size of 9 LiFi, which is because that input dimension of target and output dimension of condition in A-TCNN increase when more LiFi APs involve, and thus decrease our estimation accuracy slightly. Overall, less than 3\% performance loss is negligible for 9 LiFi case compared with dataset results. Based on this, we replace FL-based dataset method with A-TCNN and then compare the wireless network metrics versus various benchmarks in the next subsection.

\subsection{Wireless Network Metrics}
In this subsection, we evaluate two wireless network metrics: achievable throughput and Jain's fairness from the following three aspects: user number, user distribution and required data rate.

\begin{figure}[t]
\centering
\includegraphics[height=7.5cm,width=9cm]{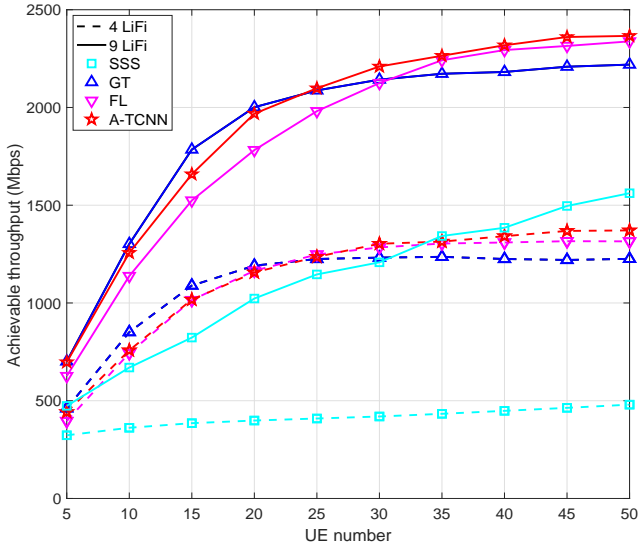}
\caption{Achievable throughput versus user number for the proposed A-TCNN compared with benchmarks.}
\label{Fig:Throughput_versus_UEnum}
\end{figure}

\subsubsection{Effect of User Number} As shown in Fig. \ref{Fig:Throughput_versus_UEnum}, we consider two general networks having size of 4 LiFi and 9 LiFi with respect to adaptive user number, where $\bar{R}$ is fixed as 100 and 200 Mbps for 4 LiFi and 9 LiFi. Note for the A-TCNN method, the throughput is measured based on the estimated output in the trained network which reaches convergence at epoch times of 20. All results are averaged using Monte-Carlo simulation. When more users are associated with APs, they compete with each other and explore more capacity resources in HLWNets, bringing the increase of achievable throughput shown in Fig. \ref{Fig:Throughput_versus_UEnum}. However, competition between users stops when it approaches the network upper bound, which is validated by figure in which the achievable throughput curves become saturating when user number increases to 50 for 4 and 9 LiFi AP cases. Specifically, the A-TCNN performs best compared with benchmarks when user number is bigger than 25 with throughput improvement about 100 Mbps, and it also remains better than FL method when user number is smaller than 25. For GT-based benchmark, it performs best when user number is less than 25, while SSS benchmark is worst as it selects AP with consideration of separate SNR and without consideration of competition and data rate requirements among users. In total, network with 9 LiFi size can provide bigger throughput (about 1000 Mbps improvement) than 4 LiFi since it involves more APs as data transmitters.

\begin{figure}[t]
\centering
\includegraphics[height=7.5cm,width=9cm]{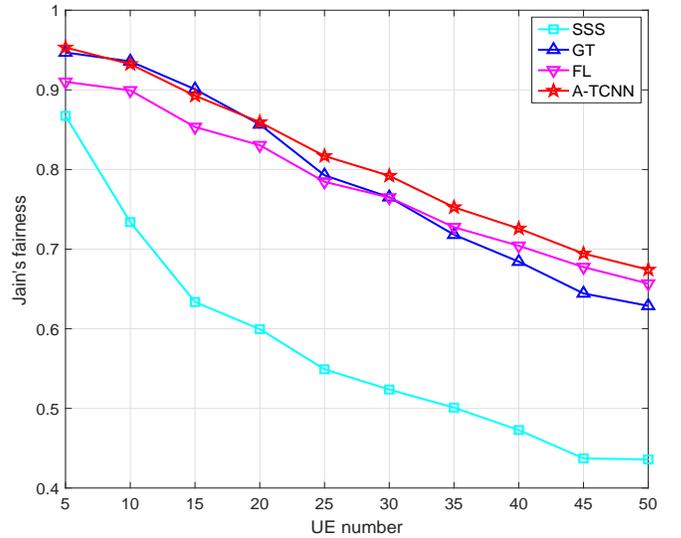}
\caption{Comparisons of Jain’s fairness index versus user number.}
\label{Fig:Fairness_UEnum}
\end{figure}

Fig. \ref{Fig:Fairness_UEnum} depicts the Jain's fairness index versus adaptive user number, where network size is chosen as 9 LiFi AP for the sake of simplicity and $\bar{R}$ is also 200 Mbps. In general, Jain's fairness drops steadily when user number becomes larger as the competition among users degrades the satisfaction values of users due to the upper bound of capacity is fixed in HLWNets. Given a small number of user, e.g. 5, the A-TCNN achieves the same fairness 0.95 as GT and FL benchmarks. For larger user number, A-TCNN performs better than benchmarks and shows 0.55 fairness compared with 0.53 fairness of GT and FL benchmarks. However, when user number is less than 25, GT method has a bigger fairness which is similar with the throughput conclusion observed in Fig. \ref{Fig:Throughput_versus_UEnum}.

\subsubsection{Effect of User Distribution}

In practical, users are not usually distributed randomly and sometimes there are a number of hot spots which serve multiple users at the same time, which is named as cluster. In Fig. \ref{Fig:Throughput_versus_UEnum}, FL method can approach the performance of A-TCNN when user number is 50. This is mainly because under the random distribution, the network upper bound has been reached with the increase of user number. As long as user number is big enough, each LiFi or WiFi AP would be overloaded, which does not show that this method owns its algorithm-wise advantage in nature.

\begin{figure}[t]
\centering
\includegraphics[height=7.5cm,width=9cm]{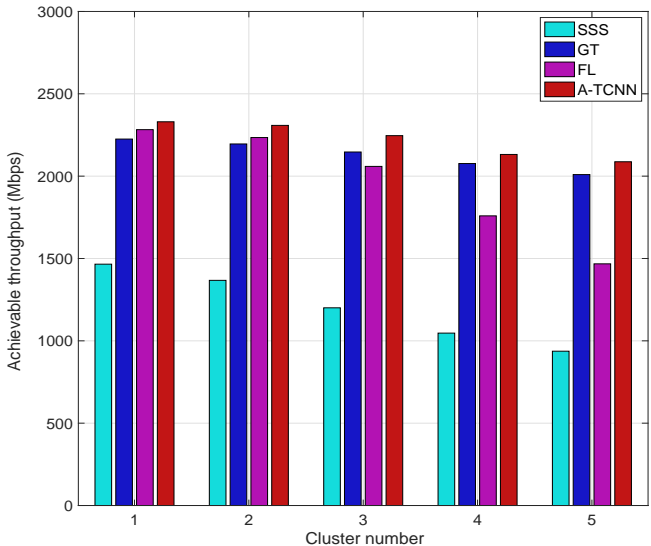}
\caption{Comparison of achievable throughput for A-TCNN versus benchmarks under effect of clustering distribution, where we consider 9 LiFi APs and 50 users for presentation.}
\label{Fig:Cluster}
\end{figure}

To verify this point and take users distribution into account, Fig. \ref{Fig:Cluster} presents the comparison of achievable throughput versus cluster number, where we assume that connected users in each cluster are subject to bivariate normal distribution, shown as $\mathbf{X} \sim \mathcal{N}_2(0, \mathbf{\Sigma})$. Here $\mathbf{\Sigma}$ denotes the covariance matrix given as [0.25 0; 0 0.25], and the number of connected users for each cluster is set as 10. As can be seen in Fig. \ref{Fig:Cluster}, the effect of clustering distribution significantly degrades the performance of SSS and FL benchmarks, while GT and proposed A-TCNN have a good tolerance ability in this scenario. When cluster number is 5, SSS and FL methods show about 38.7 and 36.5\% network throughput loss, while this percentages are only about 10.7 and 10.6 for GT and A-TCNN methods.

\subsubsection{Effect of Data Rate Requirement}
\begin{figure}[t]
\centering
\includegraphics[width=9cm]{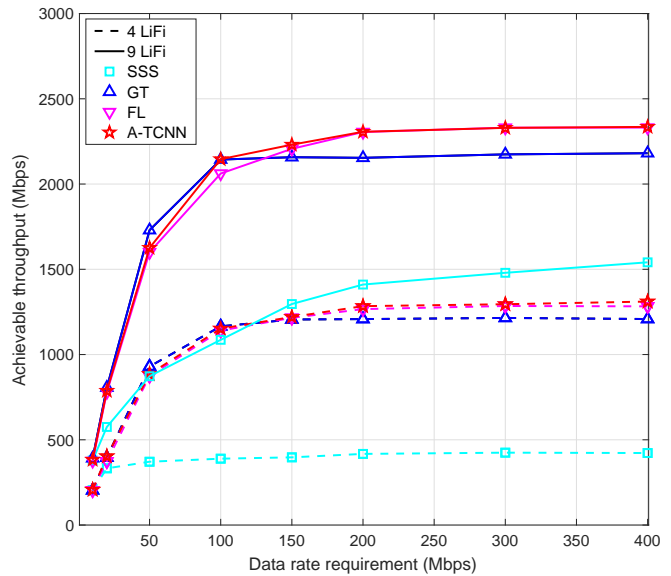}
\caption{Achievable throughput versus data rate requirement for the proposed A-TCNN load balancing method compared with benchmarks.}
\label{Fig:Throughput_versus_Rb}
\end{figure}

In Fig. \ref{Fig:Throughput_versus_Rb}, we fix user number as 20 and 40 for 4 and 9 LiFi APs respectively, and compare the achievable throughput for A-TCNN versus benchmarks with respect to the effect of data rate requirement from 10 to 400 Mbps. For 4 LiFi APs scenario, A-TCNN method improves throughput from about 200 Mbps given $\bar{R}$ of 10 Mbps to 1300 Mbps given $\bar{R}$ of 400 Mbps, where it meets saturation when $\bar{R}$ is set as about 150 Mbps. Before this threshold, A-TCNN performs slightly worse than GT based benchmark, which is negligible, while the achieved throughput gain is about 100 Mbps when $\bar{R}$ is equal to 400 Mbps. When network size is configured as 9 LiFi APs, the same tendency between benchmarks can be seen from Fig. \ref{Fig:Throughput_versus_Rb}, which illustrates that the proposed A-TCNN also has a good adaption ability to different hybrid network size. Among these benchmarks, SSS performs worst which is consistent with conclusions in Fig. \ref{Fig:Throughput_versus_UEnum} and Fig. \ref{Fig:Throughput_versus_UEnum}. For simplicity, we only present the achievable throughput metric here considering the effect of data rate requirement and no need for fairness metric presentation.

In summary, we prove that A-TCNN can perform well in terms of network throughput and fairness. It is worth noting that although GT based benchmark shows a similar performance to A-TCNN, it is a distributed algorithm that needs excessive iterations to reach the Nash equilibrium.

\begin{table}[t]
\renewcommand{\arraystretch}{1.1}
\setlength\tabcolsep{1.5 mm}
\centering
\caption{Analysis of Big-O Complexity.}
\begin{tabular}{|c|c|c|} 
\hline
\textbf{Methods} & \multicolumn{2}{c|}{\textbf{Big-O Complexity}}  \\ 
\hline
SSS  & \multicolumn{2}{c|}{$\mathcal{O}({N_{\rm{a}}}{N_{\rm{u}}})$}  \\ 
\hline
GT [\ref{Ref: Y. Wang LB Game TWC}] & \multicolumn{2}{c|}{$\mathcal{O}({N_{\rm{a}}}N_{\rm{u}}{I})$}  \\ 
\hline
FL [\ref{Ref: X. Wu AP Selection TCOM}] & \multicolumn{2}{c|}{$\mathcal{O}(N_{\rm{u}}{K_1}{K_2})$} \\ 
\hline
FL-OPT [\ref{Ref: Globecom Conference}] & \multicolumn{2}{c|}{$\mathcal{O}({N_{\rm{u}}}{K_1}{K_2}) + \mathcal{O}({N_{\rm{a}}}{N_{\rm{u}}})$} \\ 
\hline
\multicolumn{1}{|c|}{\multirow{3}{*}{A-TCNN}} & Collection &  $\mathcal{O}({N_{\rm{a}}}{N_{\rm{u}}}NT)$         \\ 
\cline{2-3}
\multicolumn{1}{|c|}{}   & Training  &  $\mathcal{O}(N_{\rm{epoch}}K_3)$  \\ 
\cline{2-3}
\multicolumn{1}{|c|}{}  & Implementation & \multicolumn{1}{c|}{$\mathcal{O}({N_{\rm{a}}}{N_{\rm{u}}}K_4 + K_5)$}  \\
\hline
\end{tabular}\label{Table: Computational Complexity}
\end{table}

\subsection{Computational Complexity}
Table \ref{Table: Computational Complexity} summarizes the Big-O complexity of the proposed method and benchmarks. With the SSS method, each user selects the AP that provides the highest SNR, leading to a computational complexity of $\mathcal{O}({N_{\rm{a}}}{N_{\rm{u}}})$. For the GT method, the Big-O complexity is estimated to be $\mathcal{O}({N_{\rm{a}}}{{{N_{\rm{u}}}}}{I})$ [\ref{Ref: Y. Wang LB Game TWC}], where $I$ denotes the number of iterations required, which turns bigger with the increase of ${N_{\rm{u}}}$. The computational complexity of the FL method depends on the number of inputs $K_{1}$ and the number of rules $K_{2}$ in the fuzzy logic. The Big-O complexity of this method can be approximated to be $\mathcal{O}({N_{\rm{u}}}{K_1}{K_2})$ [\ref{Ref: X. Wu AP Selection TCOM}]. With respect to the FL-OPT method, the part of processing fuzzy rules demands a Big-O complexity similar to that of the FL method, and the part of decision making requires $\mathcal{O}({N_{\rm{a}}}{N_{\rm{u}}})$. The A-TCNN method consists of three components: dataset collection, training, and real-time implementation. Firstly, the complexity of dataset collection is linear with the collected batch number $N$ and batch size $T$, which can be expressed as $\mathcal{O}({N_{\rm{a}}}{N_{\rm{u}}}NT)$. Secondly, training process takes several epochs to reach convergence. In each epoch, the complexity of forward and backward propagation can be denoted as $K_3$ as it is difficult to formulate. The complexity of training process can be expressed as $\mathcal{O}(N_{\rm{epoch}}K_3)$. Thirdly, in the real-time implementation of A-TCNN, complexity can be roughly given as $\mathcal{O}({N_{\rm{a}}}{N_{\rm{u}}}K_4 + K_5)$, in which $\mathcal{O}(K_4)$ represents the complexity of the matrix addition and multiplication in an FC layer, $\mathcal{O}(K_5)$ means the complexity of other operations in the TCNN, including BN and activation functions. Note the BN operation plays a dominant role in the complexity of $\mathcal{O}(K_5)$. Therefore, the changes of $N_{\rm{a}}$ and $N_{\rm{u}}$ have a minor effect on the computational complexity of implementation.

Finally, we present a comparison of runtime for the proposed A-TCNN versus various benchmarks considering different network size and user number. The results are summarized in Table \ref{Table: runtime}.\footnote{Note Bog-O notation gives upper bound for the growth rate of complexity with respect to variables in the worst case. For example, for the runtime of SSS given in Table II, it is not linear with the increase of AP number. In a basic choosing maximum element in an array, the best complexity is $\mathcal{O}(1)$ while worst case is $\mathcal{O}(N)$ as it is related to whether the array is sorted. In SSS, WiFi normally has a bigger SNR than WiFi, thus the maximum value can always be chosen in the first access.} Note for a fair comparison, we convert our learning model from Python into MATLAB platform and then measure the averaged runtime for all methods on the same computer. As can be seen in Table \ref{Table: runtime}, SSS method with simplest decision rule regarding SNR shows the lowest runtime around 0.001 $\sim$ 0.005 ms among other benchmarks, while GT, FL and FL-OPT methods all cost a long runtime and all longer than 1 ms. Between them, the proposed A-TCNN only takes a tiny runtime at a level of 0.1 ms. Consider user number and LiFi AP number, advantages of A-TCNN are more outstanding. When user number turns bigger, the FL and FL-OPT methods show a linear increment of runtime and GT method costs a exponential increment of runtime, while the A-TCNN scheme presents a stable runtime around 0.1 ms as it is adaptive to different user number in principle. In a bigger number of LiFi AP, the proposed A-TCNN requires a slightly higher runtime because the input dimensions of target and condition become higher which costs an additional processing power. When more APs are involved, benchmarks consume more runtime except from SSS and FL. For example, runtime of A-TCNN is 0.0992 and 0.123 ms for 4 LiFi, 50 users and 9 LiFi, 50 users cases, which is 91, 74 times shorter than the FL benchmark, and 480, 1503 times shorter than the GT benchmark.

\begin{table}[t]
\centering
\renewcommand{\arraystretch}{1.2}
\setlength\tabcolsep{1.6 mm}
\scriptsize
\caption{Comparison of Runtime (in Milliseconds).}
\begin{tabular}{|c|c|c|c|c|c|c|}
\hline
\multicolumn{2}{|c|}{\diagbox[]{Methods}{User No.}} & 10 & 20 & 30 & 40 & 50        \\ 
\hline
\multirow{5}{*}{4 LiFi} & SSS & 0.00132 & 0.00222 & 0.00325 & 0.00426 & 0.00501    \\ 
\cline{2-7}
                  & GT [\ref{Ref: Y. Wang LB Game TWC}] & 1.74  & 6.29 & 15.5 & 29.7 & 47.7     \\  
\cline{2-7}
                  & FL [\ref{Ref: X. Wu AP Selection TCOM}] & 2.10 & 4.00 & 5.76 & 7.40 & 9.16  \\ 
\cline{2-7}
                  & FL-OPT [\ref{Ref: Globecom Conference}] & 4.70 & 9.86 & 16.7 & 25.2 & 33.1   \\        
\cline{2-7}
                  & \textbf{A-TCNN} & \textbf{0.0819} & \textbf{0.0872} & \textbf{0.0962} & \textbf{0.0989} & \textbf{0.0992}  \\
\cline{2-7}
\hline
\multirow{5}{*}{9 LiFi} &  SSS  & 0.00152 & 0.00236 & 0.00347 & 0.00454 & 0.00501   \\ 
\cline{2-7}
                  & GT [\ref{Ref: Y. Wang LB Game TWC}] & 6.34  & 27.5 & 63.1 & 104 & 185     \\    
\cline{2-7}
                  & FL [\ref{Ref: X. Wu AP Selection TCOM}] & 2.15 & 3.94 & 5.67 & 7.56 & 9.25 \\ 
\cline{2-7}
                  & FL-OPT [\ref{Ref: Globecom Conference}] & 5.62 & 11.3 & 19.0 & 28.6 & 39.9 \\ 
\cline{2-7}
                  &  \textbf{A-TCNN}  & \textbf{0.0867} & \textbf{0.101} & \textbf{0.121} & \textbf{0.122} & \textbf{0.123}  \\
\cline{2-7}
\hline
\end{tabular}\label{Table: runtime}
\end{table}

\section{Conclusion and Future Works}
In this paper, we presented a novel adaptive target-condition neural network (A-TCNN) aided structure to tackle the LB in HLWNets. This new structure enables HLWNets being adaptive to different users and AP deployment by utilizing the generalization ability of DNN. We also introduced the dataset collection method by investigating and comparing a number of benchmarks: SSS, GT, FL, and the FL-OPT which mixes the decision-making and optimization to better balance the optimality and complexity. Training and testing results show that A-TCNN has a good convergence and scalability, together with a higher accuracy than DNN model. In addition, we verified that the trained A-TCNN can solve LB issue with near-zero performance gap and unique feature of adapting to different user number. Furthermore, simulation results illustrate that our learned A-TCNN structure can provide better network metrics in terms of achievable throughout and fairness among users. Last but not least, we analysed the computational complexity and compared the runtime for A-TCNN versus various benchmarks, in which shows that A-TCNN method can save runtime up to 458 and 1388 times compared with FL and GT benchmarks, respectively. To the best of the authors' knowledge, the proposed A-TCNN is the first work that tackles LB problem with near-optimal performance and sub-millisecond runtime in HLWNets, making the system promising for meeting the latency requirement in 6G.

Looking forward, we point out a number of interesting and promising directions for future research: 1) we consider quasi-static users in this paper for sake of simplicity while it is more realistic when consider mobility, blockage and handover in HLWNets; 2) although DNN owns generalization ability to adapt to different user number, it is unknown whether DNN-aided model can handle LB in HLWNets when consider mobile and time sequential scenario. Learning from recent progresses in ML, how to introduce a more advanced and robust model into HLWNets is an open issue, such as deep RL and generative adversarial network; 3) most works in HLWNets only focus on the physical layer while the research on network layer is also necessary and challenging. Future works will be carried out considering software-defined networking wise, which involves metrics of network packets, routing process, packet loss ratio and latency.

\appendices
\section{Proof of Optimal PA Results}
Consider AP assignments has been tackled, i.e. $\chi_{i,j}$ has been chosen as 1 in a fixed stand-alone network. Assuming the satisfaction index is relaxed as $S_j = {\sum_{i \in \mathbb{S}}{{\rho_{i,j}}{C_{i,j}}}}/{{R_j}}$, (\ref{eq:Problem2}) is equivalent to
\begin{equation}\label{eq:Problem3}
\begin{array}{l}
\max\limits_{\rho_{i,j}} \;\quad\sum\limits_{j \in \mathbb{U}} {\log S_j}\\
\;\;{\rm{s}}{\rm{.t}}{\rm{. }}\quad\;\sum\limits_{j\in\mathbb{U}} {{\rho_{i,j}} \le 1}, \forall i \in \mathbb{S},\\
\quad\quad \quad\;\; S_j \le 1, \forall\; j,\\
\quad\quad \quad\;\; 0 \le {\rho _{i,j}} \le 1, \forall\; i,j.\\
\end{array}
\end{equation}
Then, the Lagrangian function can be given as:
\begin{equation}
\mathcal{L}(\rho_{i,j}, \lambda_1, \lambda_2) =  \sum\limits_{j \in \mathbb{U}} \log S_j + \lambda_1(1 - \sum\limits_{j \in \mathbb{U}}\rho_{i,j}) + \lambda_2(1 - S_j).
\end{equation}
The KKT conditions include stationarity condition, dual feasibility, complementary slackness, and primal feasibility. These four conditions are given as:

\begin{subequations}
\begin{align}
&\nabla_{\rho_{i,j}} \mathcal{L}(\rho_{i,j}, \lambda_1, \lambda_2) = 0 , \\
&\lambda_1 \geq 0,\; \lambda_2 \geq 0, \\
&\lambda_1(1 - \sum_{j \in \mathbb{U}}\rho_{i,j}) = 0, \; \lambda_2(1 - S_j) = 0, \\
&(1 - \sum_{j \in \mathbb{U}}\rho_{i,j}) \leq 0,\; 1 - S_j \leq 0,
\end{align}    
\end{subequations}

\begin{itemize}
\item When $\lambda_1 = 0$ and $\lambda_2 > 0$, it can be derived that $S_j = 1$ for $\forall j$. Only number of users and $R_j$ are small enough, constraint of $\sum_{j\in \mathbb{U}} \rho_{i,j} \leq 1$ can be always satisfied. Thus, KKT conditions can be satisfied provided $\rho_{i,j} = R_j/C_{i,j}$ for $\forall j$.

\item When $\lambda_1 > 0$ and $\lambda_2 = 0$, similarly we have $\rho_{i,j} = 1/{\sum_{j\in \mathbb{U}} \chi_{i,j}}$ and $\rho_{i,j} \leq R_j/C_{i,j}$ to meet KKT conditions. This case is reasonable when number of users and $R_j$ are big enough.

\item When $\lambda_1 = 0$ and $\lambda_2 = 0$, the stationarity condition of (17a) can not be met, thus this case is meaningless.

\item When $\lambda_1 > 0$ and $\lambda_2 > 0$, we have $\rho_{i,j} = 1/{\sum_{j\in \mathbb{U}} \chi_{i,j}}$ and $\rho_{i,j} = R_j/C_{i,j}$ for $\forall j$, which is controversial for different users. 
\end{itemize}

In total, (\ref{eq:Problem3}) has no closed-form formulation which can satisfy all KKT conditions for different network deployment. For case of user number and $R_j$ being small, $\rho_{i,j} = 1/{\sum_{j\in \mathbb{U}} \chi_{i,j}}$ means more resources are allocated to users, bringing maximum satisfaction index as $S_j = 1$. While the theoretical result of $\rho_{i,j} = R_j/C_{i,j}$ also brings satisfaction value being 1 as fewer users with small data rate requirements are easy to serve. For case of user number and $R_j$ being large, solution of $\rho_{i,j} = 1/{\sum_{j\in \mathbb{U}} \chi_{i,j}}$ can meet all KKT conditions. When user number and $R_j$ are in medium level, only part of users can be served with $S_j = 1$ so that no theoretical solution for (\ref{eq:Problem3}) can be derived. Finally, we can approximately choose $\rho_{i,j} = 1/{\sum_{j\in \mathbb{U}} \chi_{i,j}}$ as solution of PA problem.


\end{document}